\UseRawInputEncoding
\documentclass[aps,showpacs,preprint,superscriptaddress]{revtex4}
\usepackage{graphicx}
\usepackage{subfigure}
\usepackage{float}
\usepackage{amsmath}
\usepackage{amsfonts}
\usepackage{color}
\usepackage{bm}
\usepackage{bbm}
\usepackage{txfonts}
\usepackage{array}
\usepackage{amssymb}
\begin{document}
\title{Generation of $\gamma$-photons and pairs with transverse orbital angular momentum via spatiotemporal optical vortex pulse}
\author{Cui-Wen Zhang}
\affiliation{Key Laboratory of Beam Technology of the Ministry of Education, and College of Nuclear Science and Technology, Beijing Normal University, Beijing 100875, China}
\author{De-Sheng Zhang}
\affiliation{Key Laboratory of Beam Technology of the Ministry of Education, and College of Nuclear Science and Technology, Beijing Normal University, Beijing 100875, China}
\author{Bai-Song Xie \footnote{bsxie@bnu.edu.cn}}
\affiliation{Key Laboratory of Beam Technology of the Ministry of Education, and College of Nuclear Science and Technology, Beijing Normal University, Beijing 100875, China}
\affiliation{Institute of Radiation Technology, Beijing Academy of Science and Technology, Beijing 100875, China}
\date{\today}
\begin{abstract}
We present the generation of well-collimated $\gamma$-photons and pairs with extrinsic transverse orbital angular momentum (TOAM) through the head-on collision of an intense spatiotemporal optical vortex (STOV) pulse carrying intrinsic TOAM with a high-energy electron beam. It is found that the TOAM of STOV pulse remains almost unchanged, and the TOAM is conserved in the center-of-mass frame (CMF). Moreover, there exhibits duality for particles TOAM in the CMF and laboratory frame (LF) when the initial location of high-energy electron beam is different. Furthermore, the TOAM of $\gamma$-photons in the CMF increases while that of positrons decreases as the topological charge of STOV pulse increases, whereas in the LF, the TOAM of both  $\gamma$-photons and positrons decreases. And the result under the same pulse intensity is better than that under the same pulse energy. The increase in the initial energy of high-energy electrons leads to an enhancement of the TOAM for both $\gamma$-photons and positrons in both frames. $\gamma$-photons and electrons/positrons with TOAM as a new degree of freedom maybe have an extensive applications in optical communication, astrophysics and nanomaterials and so on.

\end{abstract}
\maketitle

\section{Introduction}

Orbital angular momentum (OAM) is a crucial characteristic of light. Light sources carrying OAM can be classified into spatial optical vortex (SOV) and spatiotemporal optical vortex (STOV) according to the different phase dislocation. The SOV pulse \cite{1} carrying longitudinal OAM parallel to the propagation direction has spiral wavefront phase of the form $\exp(i\ell\phi)$, where $\ell$ represents the topological charge and $\phi$ is the azimuthal angle in the transverse plane. In past decades, researchers have extensively studied the generation of SOV pulse \cite{2,vortex2019,LG2020,LG2021} and corresponding laser-matter interactions \cite{wake2014,wake2018,pra2014,HHG2015,HHG2016,HHG2017,HHG2018,HHG20211,HHG20221,raman2023,raman20231}, especially the generation of $\gamma$-photons and pairs with longitudinal OAM   \cite{SO2017,SO2018,SO2019,SO20191,SO2020,SO2021,SO2023,ray2016,ray2016x,ray2018,ray2019,ray20191,pos2022}. Photons with OAM possesses an additional degree of freedom, which can expand the information capacity for optical communication \cite{applied2007}, detect the inhomogeneities of small-scale interstellar medium, investigate OAM transfer in Kerr black holes \cite{astro2003,astro2011} and so on. Electrons with OAM have promising applications in magnetism, nanomaterials analysis and manipulation \cite{ele2010, ele2013}.

The STOV pulse carrying pure transverse orbital angular momentum (TOAM) orthogonal to the propagation direction, is polychromatic and exhibit spiral phase in the spatiotemporal domain \cite{poly2,poly3,stov2005,stov2012}. Recently, the successful generation of STOV pulses in experiments \cite{exper2016,exper2019,exper2020} has attracted the attention of researchers, and some authors have used STOV pulses to study high-harmonic generation \cite{HHG2021,HHG2022} and spin-orbital interactions \cite{PRL2021}. To our knowledge, the generation of $\gamma$-photons and pairs with TOAM has not been investigated, which could potentially expand the applications of photons and electrons/positrons with OAM.

\begin{figure}[H]
  \centering
  \includegraphics[width=0.80\textwidth]{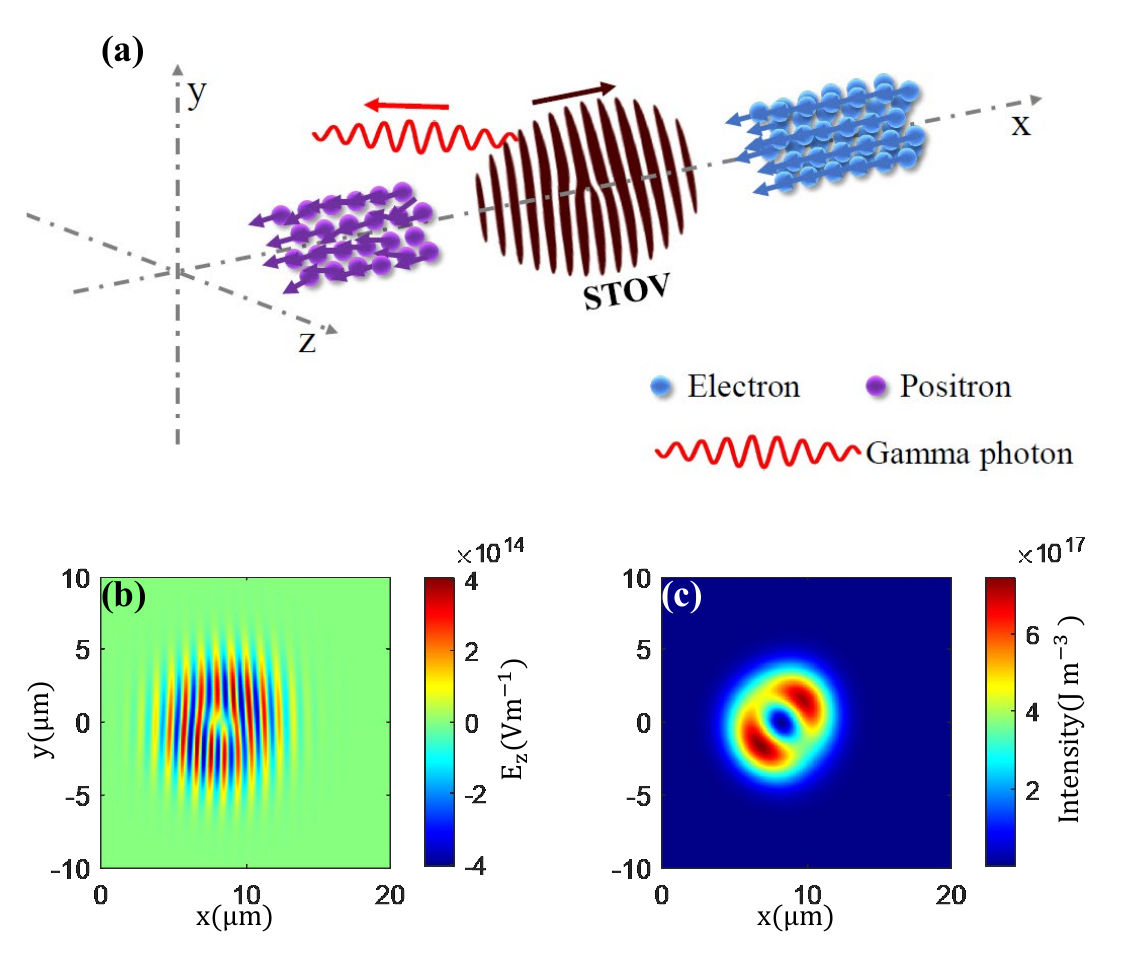}
  \caption{(color online). (a) Schematic diagram: A $z$-polarized STOV pulse collides head-on with a high-energy electron beam, whose central axis $C_a$ aligns with the $x$-axis. First, $\gamma$-photons are generated through the NCS process. Subsequently, these $\gamma$-photons further interact with the pulse and undergo the NBW process, resulting in the production of positron-electron pairs. (b) Snapshot of the electric field $E_z$ at $t=20T_0$. (c) Time-averaged energy density of the STOV pulse at $t=20T_0$.}
  \label{scheme}
\end{figure}

With the development of high-power laser technology \cite{laser2021}, it is possible to explore the quantum electrodynamics (QED) effect \cite{QED1985,QED2017}. Here we consider the generation of $\gamma$-photons and pairs with TOAM in the head-on collision of a $z$-polarized STOV pulse with an intensity of $1\times 10^{23}\rm W/cm^2$ and a 2GeV electron beam, as shown in Fig.~\ref{scheme}(a). The high-energy electron beam can be obtained by current laser wakefield accelerators \cite{MRE2024}, and the central axis $C_a$ of which in our scheme aligns with the $x$-axis. In the scheme, $\gamma$-photons are generated through the nonlinear Compton scattering (NCS) process , $e^-+n \omega_0 \rightarrow \omega_{\gamma}+e^-$, when the STOV pulse collides head-on with the high-energy electron beam. Then, the generated $\gamma$-photons annihilate into positron-electron pairs through the nonlinear Breit-Wheeler (NBW) process, $\omega_{\gamma}+n \omega_0\rightarrow e^{+} e^{-}$. The two QED processes are governed by the quantum nonlinearity parameters $\chi_e=(e\hbar/m^3_e c^4)|F_{\mu \nu}p^\nu|$ and $\chi_{\gamma}=(e\hbar^2/2m^3_e c^4)|F_{\mu \nu}k^\nu|$, respectively \cite{QED1985}, where $-e$ and $m_e$ are the charge and mass of electron, $\hbar$ is the reduced Planck constant and $c$ is the light speed in vacuum, $F_{\mu \nu}$ is the electromagnetic field tensor, $p^\nu$ and $k^{\nu}$ is the four-momentum of electron and $\gamma$-photon, respectively. When the high-energy electron beam and intense laser propagate in opposite directions, $\chi_e$ is large enough for copious $\gamma$-photons to be emitted as well as the subsequent created positrons.

We know that the TOAM of STOV pulse is intrinsic, however, in contrast, the TOAM of the initial high-energy electrons, the created $\gamma$-photons and pairs depends on the choice of coordinate frame so that they are extrinsic \cite{extrinsic2002,2015PR}. Thus, the TOAM in both the center-of-mass frame (CMF) and laboratory frame (LF) are considered in this study. It is found that the TOAM of STOV pulse remains almost unchanged, and the TOAM is conserved in the CMF. Meanwhile, we also investigate the TOAM in both frames when $C_a$ aligns with the centroid of STOV pulse and the propagation axis, respectively. For behaviors of particles TOAM in two frames, we find the duality relation between them when two $C_a$ are considered.

Furthermore, we investigate the influence of the topological charge of STOV pulse and the initial energy of high-energy electron beam in particles TOAM. It is found that, the TOAM of $\gamma$-photons  in the CMF increases while that of positrons decreases as the topological charge of STOV pulse increases, whereas the TOAM of both $\gamma$-photons and positrons in the LF declines. And the result under the same pulse intensity is better than that under the same pulse energy. The increase in the initial energy of high-energy electrons leads to an enhancement of the TOAM for both $\gamma$-photons and positrons in both frames. These results illustrate that $\gamma$-photons and pairs with large extrinsic TOAM are generated. And they have application prospects in optical communication, astrophysics and nanomaterials.

\section{Scheme setup}

The proposed scheme is performed and examined via three-dimensional (3D) QED particle-in-cell (PIC) simulations by open source code EPOCH \cite{EPOCH}, which includes the QED effect using a Monte Carlo algorithm \cite{MC}. The $z$-polarized 3D STOV pulses can be described as \cite{exper2016,exper2019},
\begin{equation}\label{field}
\begin{aligned}
{E_{z}}(x,y,z,t)=&E_{0}[(\xi/w_{\xi})^{2}+(y/w_{y})^{2}]^{|l|/2}\\
&\times \exp[-(\xi^{2}/w_{\xi}^2+y^{2}/w_{y}^2+z^2/w_{z}^2)]\\
&\times \exp[i(-l\varphi+\omega_{0}t+k_{0}z)],
\end{aligned}
\end{equation}
where $(x,y,z,t)$ are the spatial and time coordinates, $E_{0}$ is the amplitude of the laser electric field with the normalized amplitude $a_0=eE_0/m_ec\omega_0\approx270$, $\xi=x-ct$ is the spatiotemporal coupling coordinate, $w_{\xi}=w_{y}=w_{z}=w_{0}=3\mu \rm{m}$ are the spatial widths and $w_{0}$ is the spot radius. We take $l=1$, $\varphi=\rm{arctan}(y/\xi)$ is the azimuthal angle in
the spatiotemporal domain, $\omega_{0}=2\pi c/\lambda_{0}$ is the center angular frequency, $\lambda_0=1\rm\mu m$ is the wavelength and $k_{0}=\omega_{0}/c$ is the wavenumber. The pulse duration $\tau=7T_0$ with a corresponding full width at half maximum (FWHM) $\tau' \approx 11.6T_0$ and $T_0=\lambda_0/c$ is the laser period. The electron beam, located between $35\rm\mu m-40\rm\mu m$, consists of $1\times 10^{10}$ electrons with initial energy $\varepsilon_{e^-}=2\rm{GeV}$, and has a transverse spatial Gaussian distribution with the  standard deviation $\sigma=0.6\rm\mu m$, and $C_a$ aligns with the $x$-axis. The dimensions of the simulation box are $x\times y\times z=40\rm\mu m\times 20\rm\mu m\times 20\rm\mu m$ with cell $800\times 300\times 300$, and each cell contains 8 macroelectrons.

\begin{figure}
  \centering
  \includegraphics[width=0.75\textwidth]{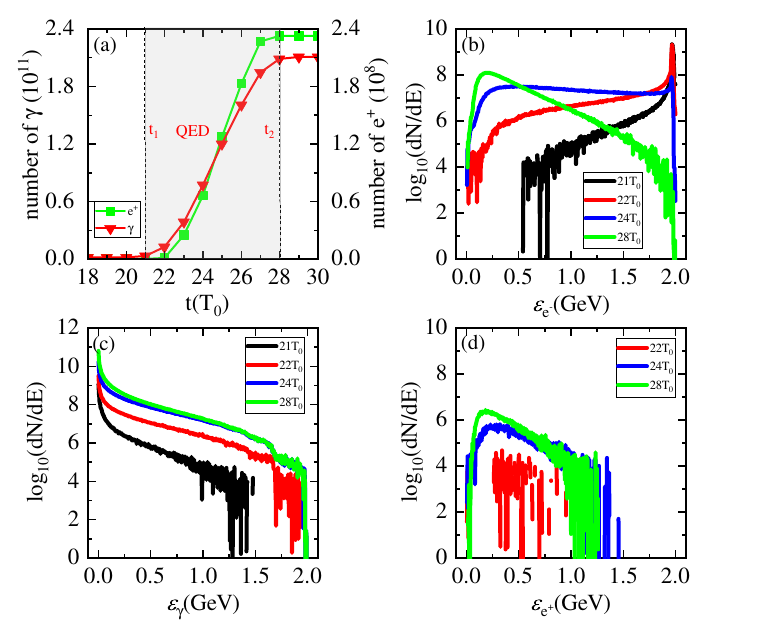}
  \caption{(color online). (a) The time evolution of the number of $\gamma$-photons and positrons. (a), (b) and (c) presents the energy spectrum of high-energy electrons, $\gamma$-photons and positrons at different times, respectively.}
  \label{spectrum}
\end{figure}

The snapshots of electric field $E_z$ and time-averaged energy density for the incident STOV pulse at $20T_0$ are depicted in Figs.~\ref{scheme}(b) and ~\ref{scheme}(c), respectively. From Fig.~\ref{scheme}(b), it can be observed that, at the centroid ($x_0=8.30\rm\mu m$, $y_0=0.078\rm\mu m$) of the STOV pulse, the electric field $E_z$ exhibits a fork-like shape, illustrating the polychromatic characteristic of STOV pulse. Fig.~\ref{scheme}(c) indicates that the intensity distribution of the STOV pulse forms a donut shape on the $xy$ plane. The time-averaged energy density can be described as $I=(\varepsilon_0|\mathbf{E}|^{2}+|\mathbf{B}|^{2}/\mu_{0})/4$ \cite{1999}, where $\varepsilon_0$ and $\mu_0$ are the dielectric constant and permeability of vacuum, respectively. Here and in the following, $\mathbf{E}$ and $\mathbf{B}$ are complex electromagnetic field, which contain envelope and phase information \cite{2019JCP}.

\section{Generation of $\gamma$-photons and pairs}

We would like to analyze the physical process since $18T_0$ when the main pulse completely enters the simulation region. Fig.~\ref{spectrum}(a) illustrates the time evolution of the number of $\gamma$-photons and positrons. The shaded area indicates that the QED process starts at $t_1=21T_0$ and ends at $t_2=28T_0$. During the process, the number of $\gamma$-photons and positrons increases and reaches a maximum of $2.1\times 10^{11}$ and $2.3\times 10^{8}$, respectively. The Figs.~\ref{spectrum}(b)-~\ref{spectrum}(d) show the energy spectrum of high-energy electrons, $\gamma$-photons and positrons at different times. It can be observed that the number of high-energy electrons with an energy of 2GeV gradually decreases, while the cutoff energy for $\gamma$-photons and positrons increases to $2\rm GeV$ and $1.5\rm GeV$, respectively. This is because that the energy of high-energy electrons is transferred to $\gamma$-photons and pairs during the QED process \cite{2017PRA}.

\begin{figure}[H]
  \centering
  \includegraphics[width=0.9\textwidth]{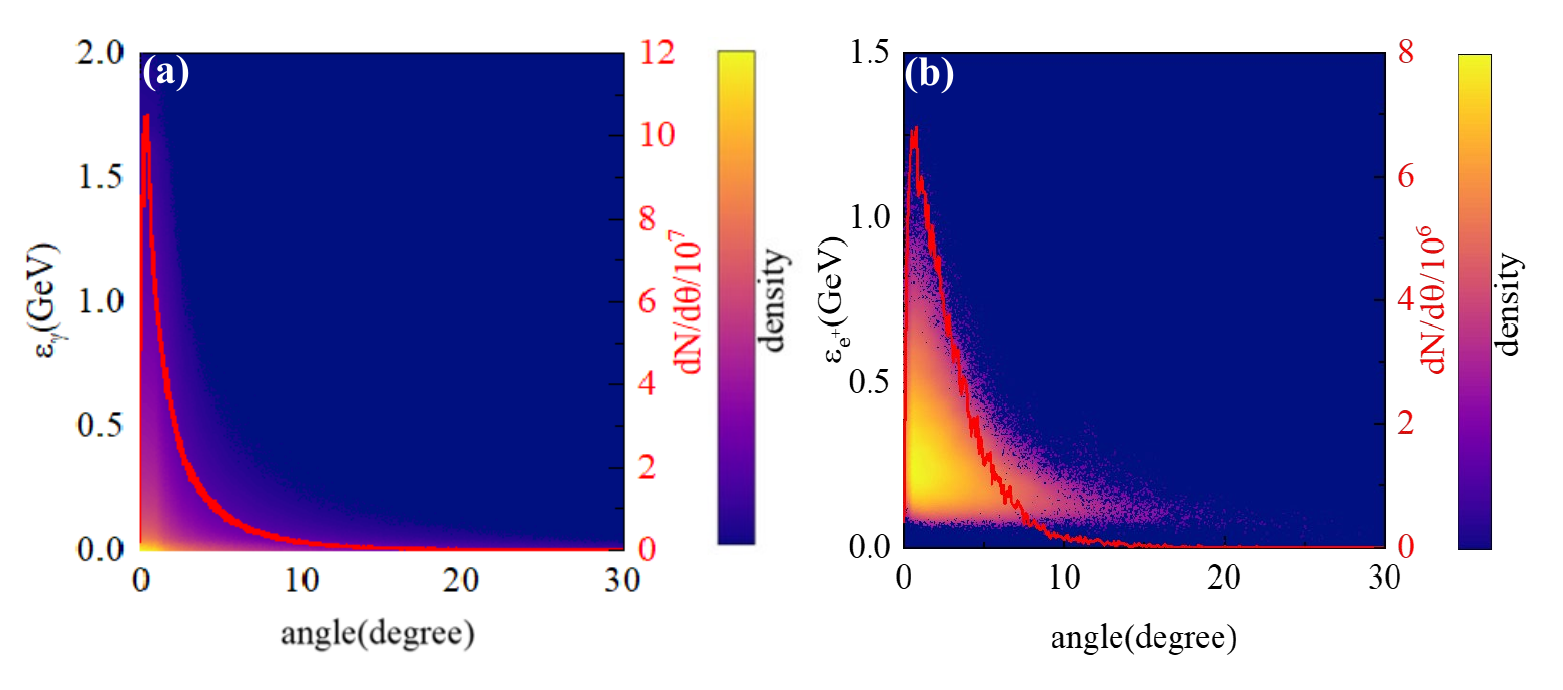}
  \caption{(color online). The polar-energy distribution of $\gamma$-photons [(a)], and positrons [(b)] at $28T_0$. The red line exhibits polar-number distribution.}
  \label{polar}
\end{figure}

The polar angle of particle can be calculated by $\arcsin(\sqrt{p^2_\bot/p^2})$ for $p_x\leq0$ and $\pi-\arcsin(\sqrt{p^2_\bot/p^2}$ for $p_x>0$, where $p^2_\bot=p^2_y+p^2_z$, $p^2=p^2_\bot+p^2_x$, and $p$ is the momentum of particle. The polar-energy distribution of $\gamma$-photons and positrons at $28T_0$ is depicted in Fig.~\ref{polar}. From Fig.~\ref{polar}(a), it can be observed that approximately $10^8$ $\gamma$-photons with an energy of $10\rm{MeV}$ are concentrated around $0.4^{\circ}$ ($\sim7\rm{mrad}$). The brightness of $\gamma$-photons at $10\rm{MeV}$ is about $5\times10^{23}$ photons ${\rm/(s\cdot mm^2\cdot mrad^2\cdot 0.1\%BW})$. Fig.~\ref{polar}(b) illustrates that around $3\times10^6$ positrons with an energy of $180\rm{MeV}$ are concentrated around $0.8^{\circ}$ ($\sim14\rm{mrad}$), exhibiting a brightness of positrons at this energy level to be about $5.8\times10^{22}$ positrons $ {\rm/(s\cdot mm^2\cdot mrad^2\cdot 0.1\%BW})$. These findings indicate that well-collimated $\gamma$-photons and positrons are generated.

\section{TOAM of STOV pulse and particles}

Here, we consider the variation in TOAM. The canonical momentum density which appears in canonical Noether conservation laws in electromagnetic field theory is used to describe the momentum of the STOV pulse. It can be expressed as \cite{2015PR}
\begin{equation}\label{momentum}
\begin{aligned}
\mathbf{P}=\rm{Im}(\varepsilon_0|\mathbf{E}^*\cdot (\nabla)\mathbf{E}+|\mathbf{B}|^{2}/\mu_{0})/4\omega,
\end{aligned}
\end{equation}
where $\textbf{X}\cdot\textbf{(Y)Z}\equiv\sum_i X_i\textbf{Y}Z_i$, indicates that the canonical momentum density depends on the local gradient of the phase of the electromagnetic field. And the TOAM of the STOV pulse with the propagational term can be calculated by \cite{PRL2021}
\begin{equation}\label{TOAM}
{L^{laser}_z}=\sum[[(x-x_0)P_y-(y-y_0)P_x]-(y-y_0)I/c)]=nl\hbar,
\end{equation}
where $(x,y)$ represent the position of cell, and $n$ is the photon number of the pulse.

The TOAM of particles is extrinsic, so we would consider the TOAM of particles in both the CMF and LF. Coincidentally, the origin of CMF coincides with the centroid ($x_0, y_0$) of STOV pulse. Thus, the TOAM of particles within the CMF can be calculated as $L^{particle}_z=\sum[(x-x_0)p_y-(y-y_0)p_x]$. In the LF, the TOAM of particles is calculated with $L^{particle}_z=\sum(xp_y-yp_x)$. In addition, we move the high-energy electron beam upwards to align $C_a$ with $y=y_0$, and investigate particles TOAM in both frames. Next, we will conduct a detailed analysis of these cases.

\subsection{$C_a$ aligns with the $x$-axis}

\begin{figure}[t]
\centering
\subfigure{
\includegraphics[trim=0.5cm 0.5cm 0.5cm 0.5cm, clip, width=8cm,height = 5cm]{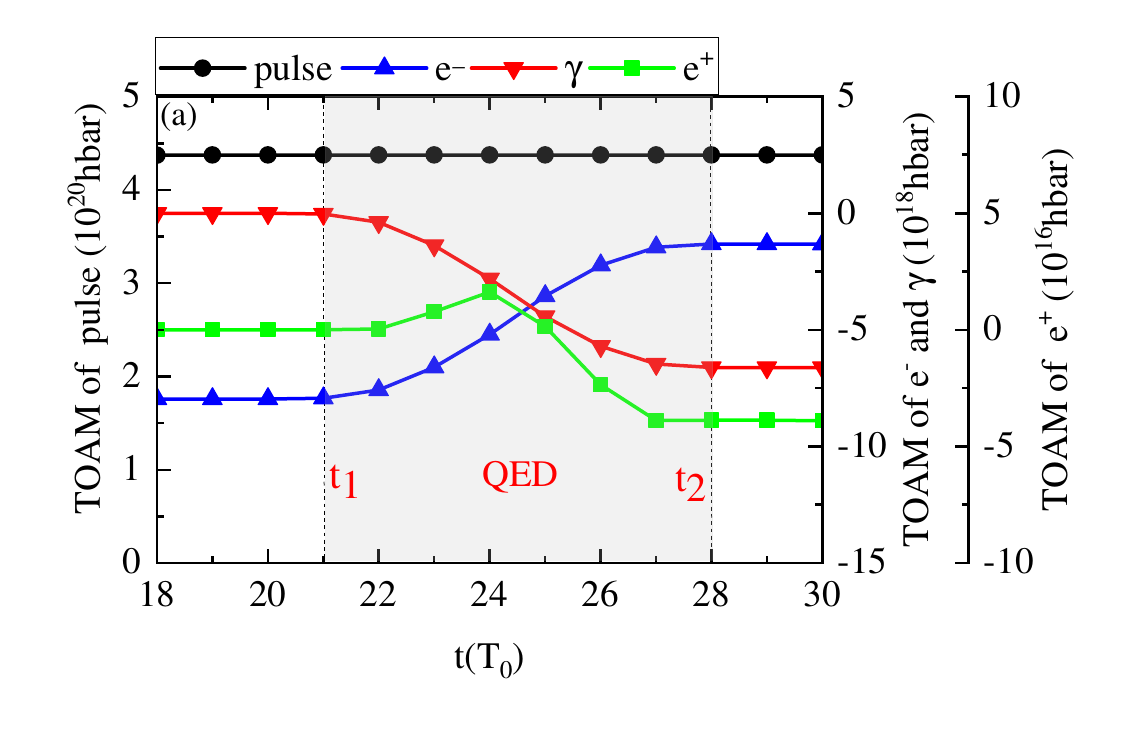}}
\subfigure{
\includegraphics[width=8cm,height = 5cm]{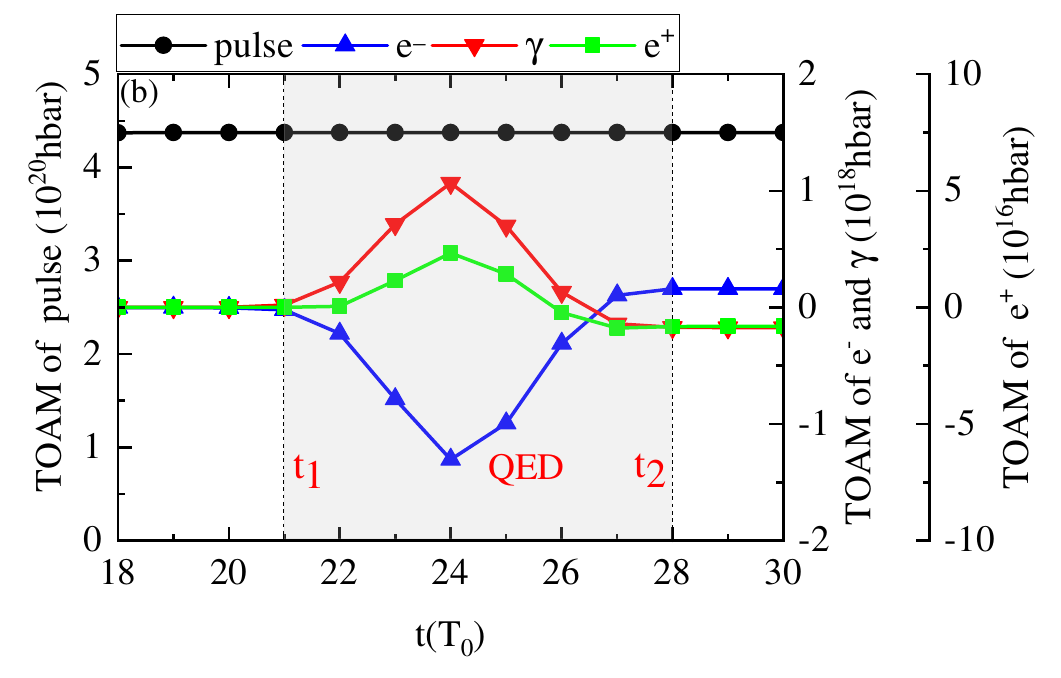}}
\caption{(color online). The time evolution of the TOAM of STOV pulse, high-energy electron beam, $\gamma$-photons and positrons in the CMF [(a)] and LF [(b)]. The STOV pulse with intensity of $10^{23}\rm W/cm^2$ collides head-on with 2GeV electron beam, and $C_a$ aligns with the $x$-axis.}
\label{x}
\end{figure}

In this section, we investigate the TOAM for pulse and particles in both frames when $C_a$ aligns with the $x$-axis. The Fig.~\ref{x}(a) shows the time evolution of the TOAM of STOV pulse, high-energy electron beam, $\gamma$-photons and positrons in the CMF. The TOAM of the STOV pulse is $4.37\times 10^{20}\hbar$, and due to the utilization of the CMF, the high-energy electron beam initially possesses a negative TOAM of $-7.97\times 10^{18}\hbar$. During the QED process, the TOAM of the STOV pulse remains almost unchanged (further explanation will be provided later), while that of the high-energy electron beam decreases to $-1.33\times10^{18}\hbar$, and that of the $\gamma$-photons and positrons reaches a maximum of $-6.6\times10^{18}\hbar$ and $-3.88\times10^{16}\hbar$, respectively. The magnitude of particles TOAM is close to that of the longitudinal OAM \cite{SO20191,SO2021,SO2023,ray2016}. It is found that the change in the TOAM of high-energy electron beam, $6.64\times10^{18}\hbar$,  is consistent with the sum of the change in TOAM for $\gamma$-photons and pairs. This is because the TOAM of the pulse-particle system can be recognized as being calculated in the frame of "torque-free", which means that the TOAM is conserved in the CMF.

The Fig.~\ref{x}(b) shows the time evolution of the TOAM of STOV pulse, high-energy electron beam, $\gamma$-photons and positrons in the LF. It indicates that, in the QED process, the TOAM of the STOV pulse still remains unchanged while the TOAM of high-energy electron beam, $\gamma$-photons and positrons reaches maximum of $-1.30\times10^{18}\hbar$, $1.06\times10^{18}\hbar$ and $2.32\times10^{16}\hbar$ at $24T_0$, and the FWHM of particle TOAM is approximately $10\rm{fs}$. The change in the TOAM of high-energy electron beam, $1.30\times10^{18}\hbar$, is not equal to the sum of the change in TOAM for $\gamma$-photons and positrons, which is $1.08\times10^{18}\hbar$. Therefore, the TOAM is not conserved in the LF.

These findings suggest that, in the QED process, the STOV pulse induces extrinsic TOAM transfer between particles while keeping its own intrinsic TOAM unchanged. Additionally, it is found that there exists a disparity in the variations of particles TOAM between two frames caused by the deviation $y_0$, and TOAM is conserved in the CMF.

\subsection{$C_a$ aligns with $y=y_0$}

\begin{figure}[t]
\centering
\subfigure{
\includegraphics[width=8cm,height = 5.1cm]{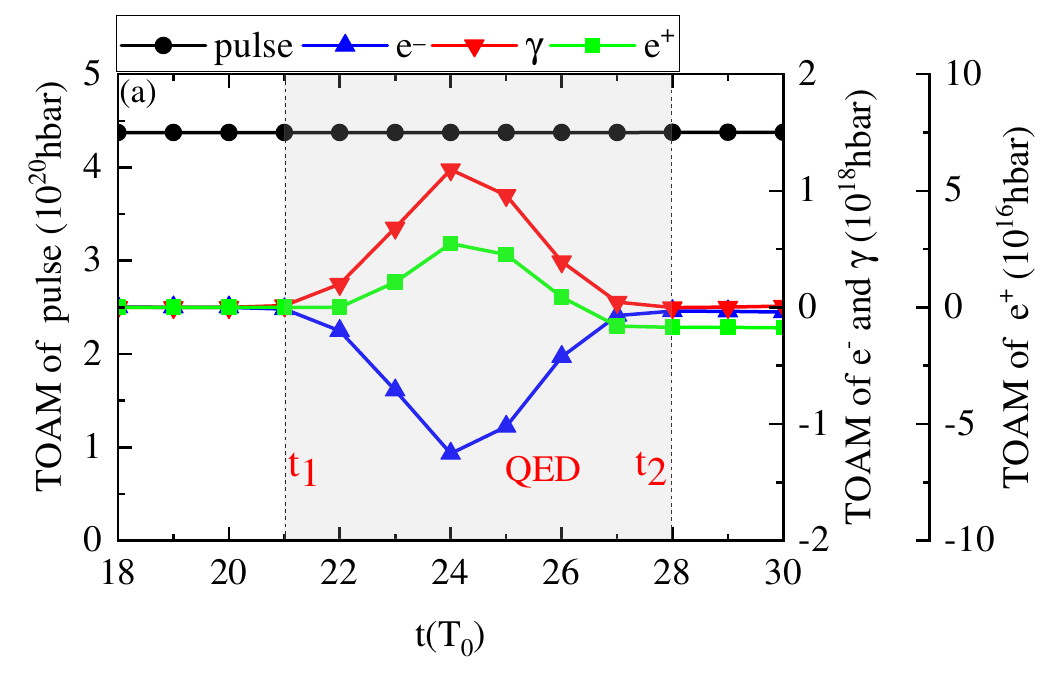}}
\subfigure{
\includegraphics[trim=0cm 0.1cm 0cm 0cm, clip, width=8cm,height = 5cm]{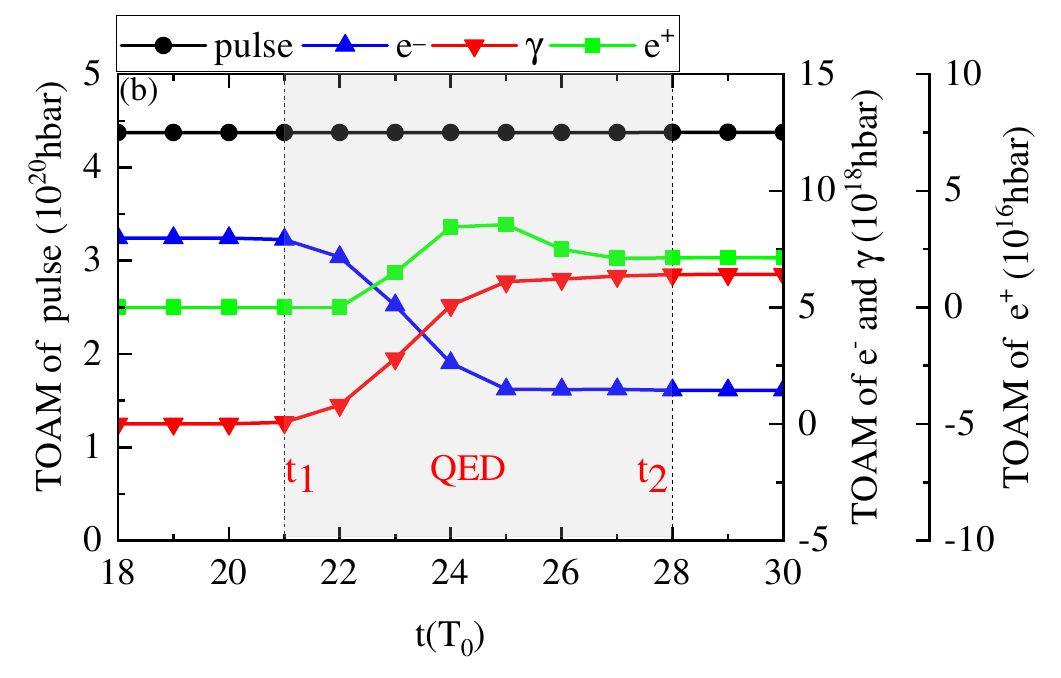}}
\caption{(color online). The time evolution of the TOAM of STOV pulse, high-energy electron beam, $\gamma$-photons and positrons in the CMF [(a)] and LF [(b)]. The STOV pulse with intensity of $10^{23}\rm W/cm^2$ collides head-on with 2GeV electron beam, and $C_a$ aligns with $y=y_0$.}
\label{y0}
\end{figure}

When we move the high-energy electron beam upward to align $C_a$ with $y=y_0$, let us see how the TOAM of laser pulse and particles behave in both frames. In this case, the time evolution of the TOAM of STOV pulse, high-energy electron beam, $\gamma$-photons and positrons are shown in Fig.~\ref{y0}. Comparison with Fig.~\ref{x} reveals that the behavior of time evolution of TOAM in the CMF (LF) when $C_a$ aligns with $y=y_0$ ($y=0$) is consistent with that in the LF (CMF) when $C_a$ aligns with the $y=0$ ($y=y_0$). In other words, there exists a duality relation between CMF-LF in terms of the two cases of $C_a$. Besides, however, the "TOAM is conserved" only holds in the CMF, which does not matter to duality. These quantities are displayed in Table ~\ref{Table 1}.

\begin{table}[H]
\caption{The TOAM changes of high-energy electron beam, $\gamma$-photons, positrons and the sum of $\gamma$-photons and pairs in the CMF and LF when $C_a$ aligns with the $x$-axis (y=0) and $C_a$ aligns with $y=y_0$.}
\centering
\begin{ruledtabular}
\begin{tabular}{cccccc}
  cases & $e^-$ ($10^{18}\hbar$) & $\gamma$ ($10^{18}\hbar$) & $e^+$ ($10^{18}\hbar$) & $\gamma$+$(e^+e^-)$  ($10^{18}\hbar$)\\
\hline
CMF ($y=0$)   & $6.64$ & $6.61$ & $0.04$ & $6.69$ \\
LF ($y=0$)    & $1.30$ & $1.06$ & $0.02$ & $1.10$ \\
CMF ($y=y_0$) & $1.25$ & $1.18$ & $0.03$ & $1.24$ \\
LF ($y=y_0$)  & $6.54$ & $6.41$ & $0.02$ & $6.45$ \\
\end{tabular}
\end{ruledtabular}
\label{Table 1}
\end{table}

\section{Influences by parametric changes}

Due to the existed duality, we would only analyze the cases when $C_a$ aligns with the $x$-axis in the following. In order to verify the invariance of TOAM for STOV pulse in the QED process, the head-on collision of a STOV pulse with intensity of $10^{21}\rm W/cm^2$ and a $10\rm{GeV}$ electron beam is studied. Furthermore, the particles TOAM could be influenced by the parameters of the STOV pulse and high-energy electron beam, especially the topological charge $l$ and the initial energy of high-energy electrons $\varepsilon_{e^-}$. So, we also conduct research on these two parameters.

\subsection{Another set of laser intensity and initial electron energy}

\begin{figure}[H]
\centering
\subfigure{
\includegraphics[trim=0cm 0cm 0cm 0cm, clip, width=8cm,height = 5cm]{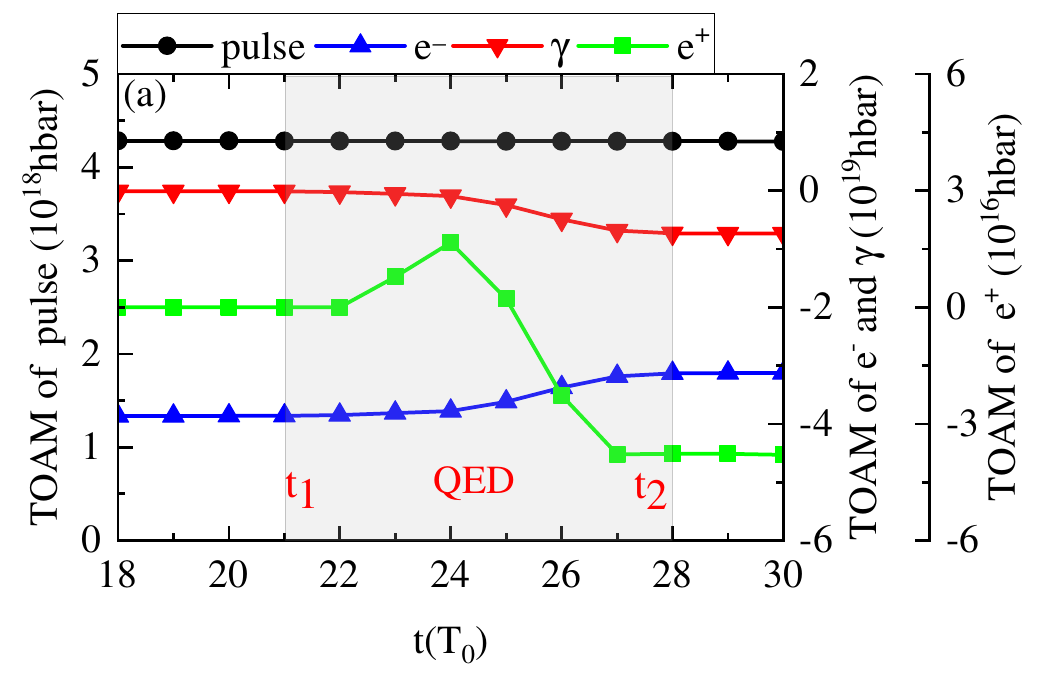}}
\subfigure{
\includegraphics[trim=0.4cm 0.25cm 0.3cm 0cm, clip, width=8cm,height = 5cm]{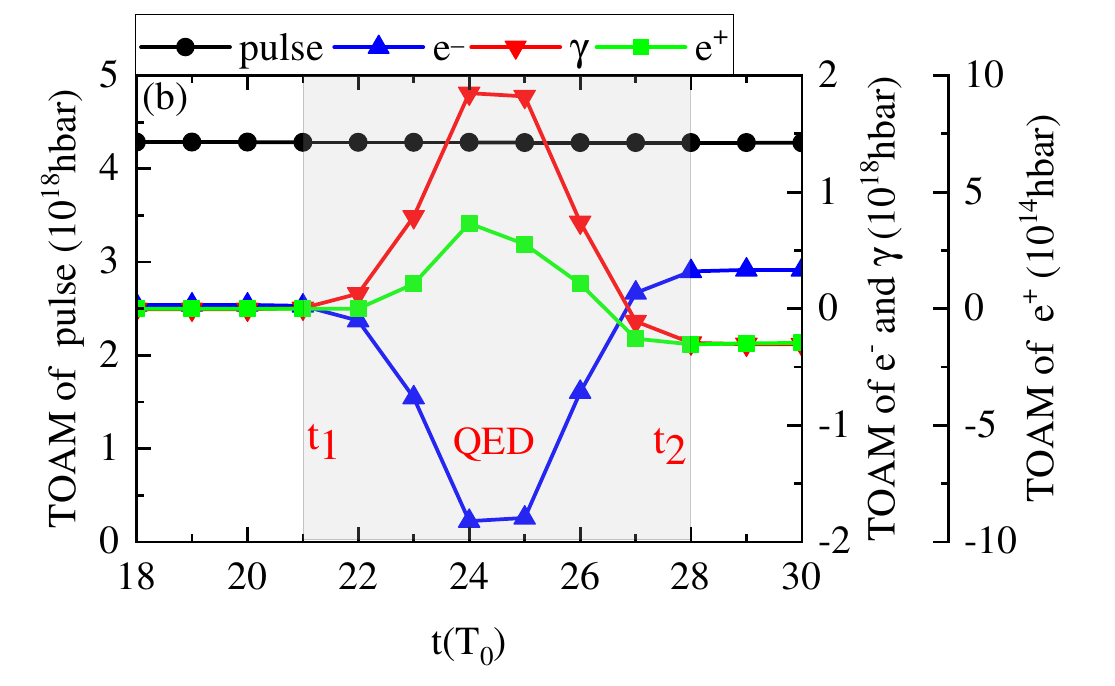}}
\caption{(color online). The time evolution of the TOAM of STOV pulse, high-energy electron beam, $\gamma$-photons and positrons in the CMF [(a)] and LF [(b)]. The STOV pulse with intensity of $10^{21}\rm W/cm^2$ collides head-on with 10GeV electron beam, and $C_a$ aligns with the $x$-axis.}
\label{2110}
\end{figure}

The head-on collision between the STOV pulse with an intensity of $10^{21}\rm W/cm^2$ and an initial $10 \rm GeV$ electron beam is simulated. The Fig.~\ref{2110} shows the time evolution of the TOAM of STOV pulse, high-energy electron beam, $\gamma$-photons and positrons in both frames.

From Fig.~\ref{2110}(a), it can be observed that the TOAM of STOV pulse still remains unchanged during the QED process with value of $4.28\times 10^{18}\hbar$ which is lower than that of $10^{23}\rm W/cm^2$ pulse. This is because the low-intensity STOV pulse has a smaller number of photons, which leads to the reduction of TOAM. The high-energy electron beam initially possesses a higher negative TOAM of $-3.86\times 10^{19}\hbar$ due to the increase of energy. At $28T_0$, the TOAM of the high-energy electron beam decreases to $-3.13\times10^{19}\hbar$, and that of the $\gamma$-photons and positrons reaches a maximum of $-7.37\times10^{18}\hbar$ and $-3.77\times10^{16}\hbar$, respectively. The change in TOAM in the CMF is $7.37\times10^{18}\hbar$, which is larger than that in the case of $10^{23}\rm{W/cm^2}-2GeV$. As can be seen from Fig.~\ref{2110}(b), the change in TOAM of high-energy electron beam, $\gamma$-photons and positrons in the LF is $-1.82\times10^{18}\hbar$, $1.85\times10^{18}\hbar$ and $3.66\times10^{14}\hbar$, respectively, and is also larger than that in the case of $10^{23}\rm{W/cm^2}-2GeV$.

The quantum parameters are lower than that in the case of $10^{23}\rm{W/cm^2}-2GeV$, and would lead to a decrease in the number of $\gamma$-photons and pairs. While the higher energy electron beam can transfer more energy to $\gamma$-photons and pairs. Thus, the influence of electron beam energy dominates here.
Furthermore, the result demonstrates that the invariance of TOAM for STOV pulse is general, and the STOV pulse stimulates the transfer of TOAM between particles due to its fork-like electric field.

\begin{figure}
  \centering
  \includegraphics[width=0.60\textwidth]{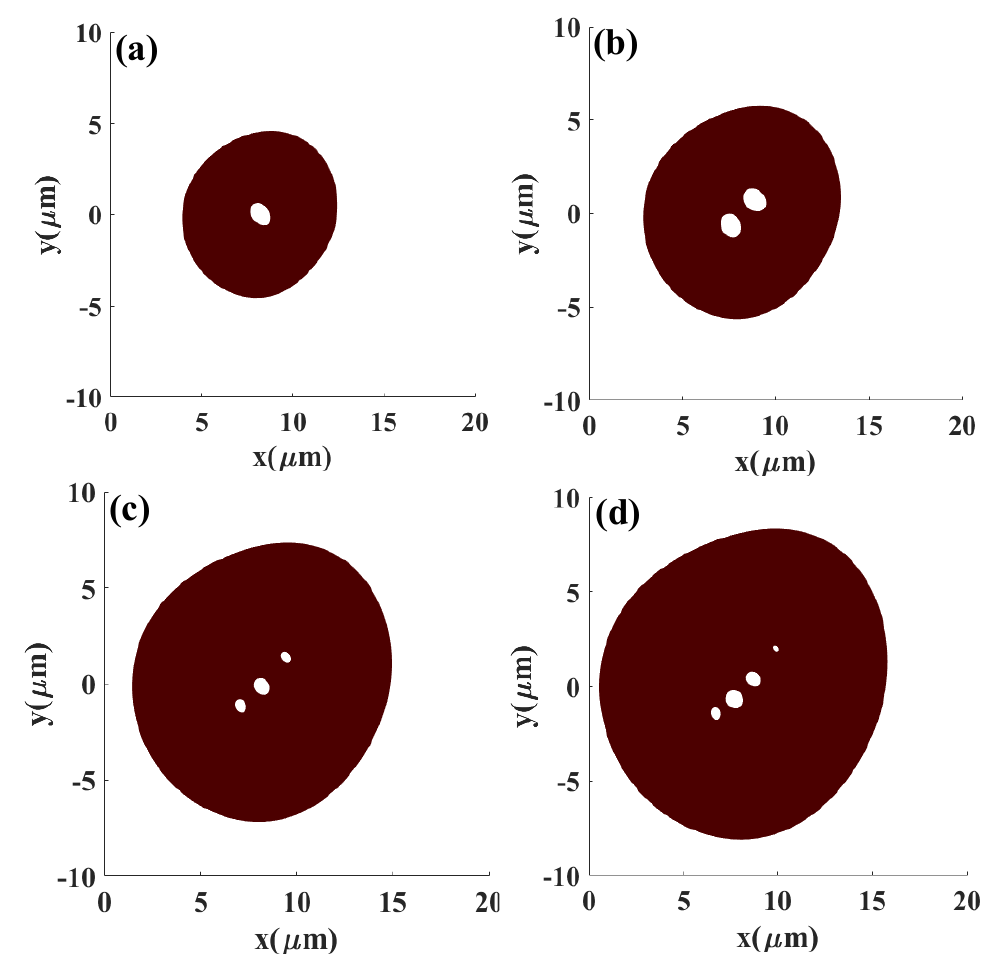}
  \caption{(color online). The iso-intensity profiles of the STOV pulse for different topological charge (a) $l=1$, (b) $l=2$, (c) $l=3$ and (d) $l=4$.}
  \label{topological}
\end{figure}

\begin{figure}
  \centering
  \includegraphics[width=0.9\textwidth]{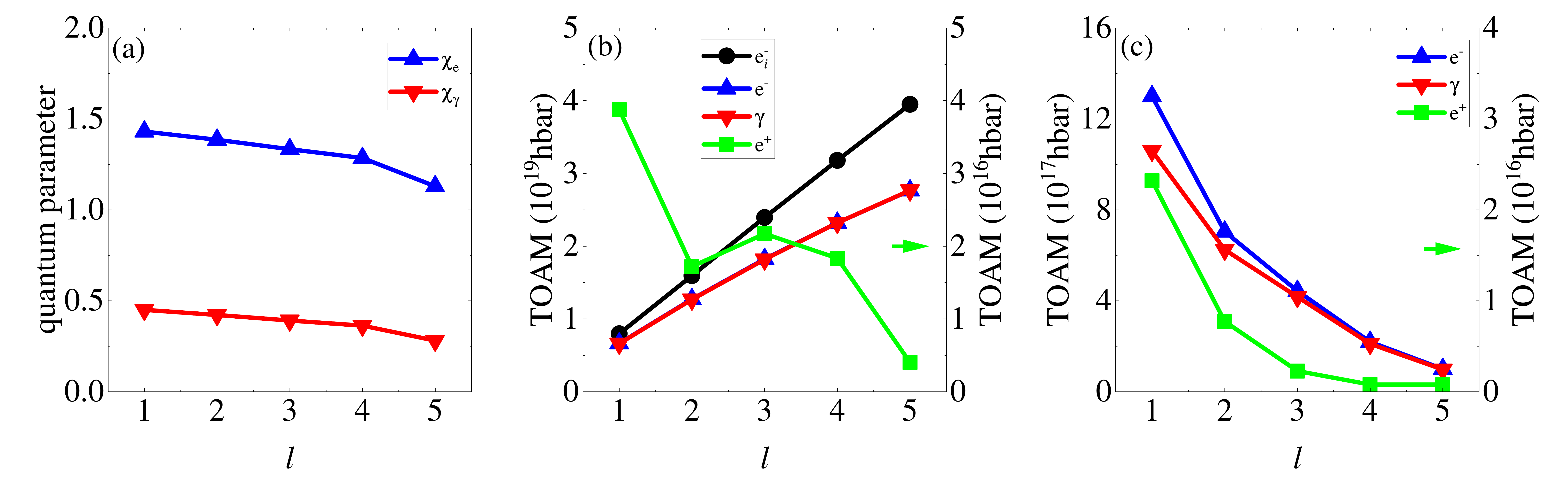}
  \caption{(color online). (a) The variation of the quantum parameters with $l$. (b) The variation of the initial TOAM of the high-energy electron beam, the TOAM of high-energy electron beam, $\gamma$-photons and positrons with $l$ in the CMF. Please note that the lines representing the high-energy electron beam and $\gamma$-photons are almost overlaping. (c) The variation of the TOAM of high-energy electron beam, $\gamma$-photons and positrons with $l$ in the LF. The green lines in (b) and (c) correspond to the right $y$-axis scale. The pulse energy is the same.}
  \label{l}
\end{figure}

\subsection{Topological charge $l$}

\begin{table}[H]
\caption{The scaling laws for the TOAM of $\gamma$-photons and positrons with respect to $l$, in the case of same pulse energy and same pulse intensity in the CMF and LF, respectively.}
\centering
\begin{ruledtabular}
\begin{tabular}{cccccc}
   cases & $\gamma$  & $e^+$\\
\hline
CMF (same pulse energy)    & $l^{0.8927}$  & $l^{-1.0296}$  \\
LF (same pulse energy)     & $l^{-1.3863}$ & $l^{-2.2842}$  \\
CMF (same pulse intensity) & $l^{1.14}$    & $l^{-0.78}$    \\
LF (same pulse intensity)  & $l^{-1.2616}$ & $l^{-2.0348}$  \\
\end{tabular}
\end{ruledtabular}
\label{Table 2}
\end{table}

Here, the changes in $l$ with values of 1, 2, 3, 4, and 5 are studied while keeping the laser energy as a same constant. Fig.~\ref{topological} illustrates the iso-intensity profiles with $l=$1, 2, 3, 4. It can be observed that the number of spatiotemporal holes is equal to the value of $l$ \cite{exper2020}. And the $y_0$ is $0.078\rm{\mu m}$, $0.157\rm{\mu m}$, $0.236\rm{\mu m}$ and $0.315\rm{\mu m}$ respectively, which increases linearly with $l$.

From Eq.~\eqref{field}, the $E_z$ decreases as $l$ increases. When $\chi_e>1$, the energy of $\gamma$-photon is $h\omega_{\gamma}\approx0.44\chi_e\gamma_e$, where $\gamma_e$ is the Lorentz factor of high-energy electron. In our studied case, there is a relation between two nonlinear quantum parameters as $\chi_{\gamma}\approx0.22\chi_e^2$ (refer to \cite{chi2008,chi2009,chi2011}). Consequently, the $\chi_e$ and $\chi_\gamma$ decreases with increasing $l$, as illustrated in Fig.~\ref{l}(a). From Fig.~\ref{l}(b), we can see that the initial TOAM of the high-energy electron beam in the CMF increases linearly with $l$, which is due to the increase of $y_0$. Moreover, the change in the TOAM of high-energy electron beam and $\gamma$-photons increases linearly with the variation of $l$, while that of the positrons decreases with $l$. Since that an increase in $y_0$ will lead to the increase in TOAM of a single particle, while a decrease in the quantum parameter will result in a decrease in the number of particle. Thus, we conclude that for high-energy electron beam and $\gamma$-photons, the influence of $y_0$ dominates, while for positrons, the influence of the quantum parameter dominates. Fig.~\ref{l}(c) shows the change in the TOAM of high-energy electron beam, $\gamma$-photons and positrons in the LF and it is found that all of them decreases as $l$ increases. This is because the TOAM of particles in the LF is irrelevant with $y_0$, and the quantum parameters dominate to all of them.

The scaling laws for the TOAM of $\gamma$-photons and positrons with respect to $l$ in two frames of CMF and LF under the same pulse energy or/and same intensity are presented in Table ~\ref{Table 2}. The scaling laws under the same pulse intensity are obtained by calibrating the quantum parameters based on those at $l=1$ so that the exponentials of the scaling law in this case would be better than those under the same pulse energy.

\subsection{Initial energy of high-energy electron beam}

\begin{figure}
  \centering
  \includegraphics[width=0.9\textwidth]{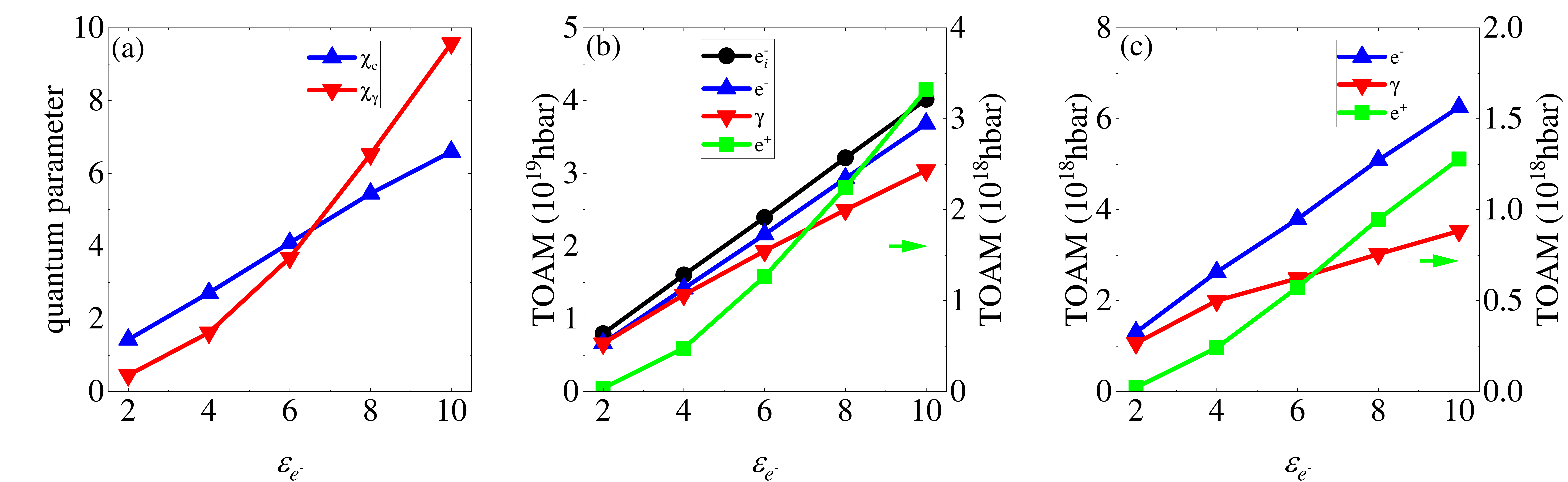}
  \caption{(color online). (a)The variation of the quantum parameters with $\varepsilon_{e^-}$. (b)The variation of the initial TOAM of the high-energy electron beam, the TOAM of high-energy electron beam, $\gamma$-photons and positrons with $\varepsilon_{e^-}$ in the CMF. (c) The variation of the TOAM of high-energy electron beam, $\gamma$-photons and positrons with $\varepsilon_{e^-}$ in the LF. The green lines in (b) and (c) correspond to the right $y$-axis scale.}
  \label{energy}
\end{figure}

In this section, the changes in $\varepsilon_{e^-}$ are studied at values of $2\rm GeV$, $4\rm GeV$, $6\rm GeV$, $8\rm GeV$ and $10\rm GeV$. Fig.~\ref{energy} indicates that all the quantities in two frames increase with the increase of $\varepsilon_{e^-}$. This is because the quantum parameters increase with $\varepsilon_{e^-}$, and the higher energy electrons could transfer more energy to $\gamma$-photons and positrons, then it further leads to an increase in particles TOAM. And due to the contribution of $y_0$, the quantities in the CMF are still larger than those in the LF.

By comparing with each other, it can be found that the increase in $l$ and $\varepsilon_{e^-}$ has almost the same effect on the increase of the initial TOAM of high-energy electron beam, while they have opposite impacts on quantum parameters. The increase in $\varepsilon_{e^-}$ can also lead to a greater number of $\gamma$-photons and positrons, and both the $\gamma$-photons and positrons have higher TOAM. Furthermore, it is easier to experimentally achieve electron beams with higher energy than to construct STOV pulses of complex structure. Thus, we think that it is a more effective way via the energy increasing of the high-energy electron beam to achieve the higher TOAM for $\gamma$-photons and positrons.

\section{Conclusion}

In a summary, we investigate the generation of well-collimated $\gamma$-photons and pairs with large extrinsic TOAM via the head-on collision of an intense STOV pulse and a high-energy electron beam. It is found that the TOAM of the STOV pulse remains unchanged, while its fork-like electric field induces the extrinsic TOAM of particles (photons/pairs) in QED processes.

The comparison between the CMF and LF indicates that the TOAM in the CMF is conserved. And the comparison between $C_a$ aligns with the $x$-axis and $C_a$ aligns with the $y=y_0$ shows that there exists the duality relation in particles TOAM variation behaviors in two frames of CMF anf LF. Furthermore, the investigation in the $l$ indicates that the TOAM of $\gamma$-photons in the CMF increases while that of positrons decreases with $l$, whereas both the TOAM of $\gamma$-photons and positrons in the LF decreases. And the results under the same pulse intensity are better than those under the same pulse energy. The investigation in the $\varepsilon_{e^-}$ shows that the he TOAM for both $\gamma$-photons and positrons in both frames are enhanced when $\varepsilon_{e^-}$ is larger.

Our proposed scheme provides a method for producing $\gamma$-photons and pairs carrying TOAM, which is believed to be have some applications to fields such as optical communication, astrophysics and nanomaterials and so on.

\begin{acknowledgments}

The authors are grateful to L.G.Zhang, H.H.Fan and Z.H.Feng for stimulating discussions. This work was supported by the National Natural Science Foundation of China (NSFC) under Grant No.12375240 and No.11935008. The computation was carried out at the HSCC of the Beijing Normal University. We acknowledge the open source PIC code EPOCH.
\end{acknowledgments}

\end{document}